\documentclass[%
 aip,
 amsmath,amssymb,
 reprint,%
]{revtex4-1}

\usepackage{graphicx}
\usepackage{dcolumn}
\usepackage{bm}

\usepackage[utf8]{inputenc}
\usepackage[T1]{fontenc}
\usepackage{mathptmx}

\begin{document}

\title{High signal to noise absorption imaging of alkali atoms at moderate magnetic fields}

\author{Maurus Hans}
\email{highfieldimaging@matterwave.de}
\author{Finn Schmutte}
\author{Celia Viermann}
\author{Nikolas Liebster}
\author{Marius Sparn}
\author{Markus K. Oberthaler}
\author{Helmut Strobel}
\affiliation{Kirchhoff-Institut für Physik, Universität Heidelberg, Im Neuenheimer Feld 227, 69120 Heidelberg, Germany.}

\date{\today}

\begin{abstract}
We present an improved scheme for absorption imaging of alkali atoms at moderate magnetic fields, where the excited state is well in the Paschen-Back regime but the ground state hyperfine manifold is not. It utilizes four atomic levels to obtain an approximately closed optical cycle. With the resulting absorption of the corresponding two laser frequencies we extract the atomic column density of a $^{39}$K Bose-Einstein condensate. The scheme can be readily applied to all other alkali-like species.
\end{abstract}

\maketitle

\section{Introduction}

Absorption imaging is a standard technique for observations in quantum gas experiments which relies on resonant atom light interaction in ideally closed optical cycle schemes.\cite{ketterleMakingProbingUnderstanding1999}
At very high magnetic fields in the Paschen-Back regime these can be found for every ground state.
At moderate fields, where only the excited state is well in the Paschen-Back regime, this is only possible for the atom's stretched states with maximal or minimal magnetic quantum number. Efficient optical pumping schemes to reach these stretched states  \cite{berningerUniversalThreeFourbody2011, berningerPaper}
are not available for arbitrary initial states.
However, when using Feshbach resonances  to tune the atomic interaction strength,\cite{chinFeshbachResonancesUltracold2010} the choice of atomic states is fixed.
Recently, a scheme for fluorescence imaging has been developed that improves the single-atom detection in these states.\cite{BergschneiderImaging} It makes use of two atomic transitions in order to obtain an approximately closed four-level optical cycle. Here, we adapt this scheme to absorption imaging of dense atomic clouds.
 
\begin{figure}
    \centering
    \includegraphics{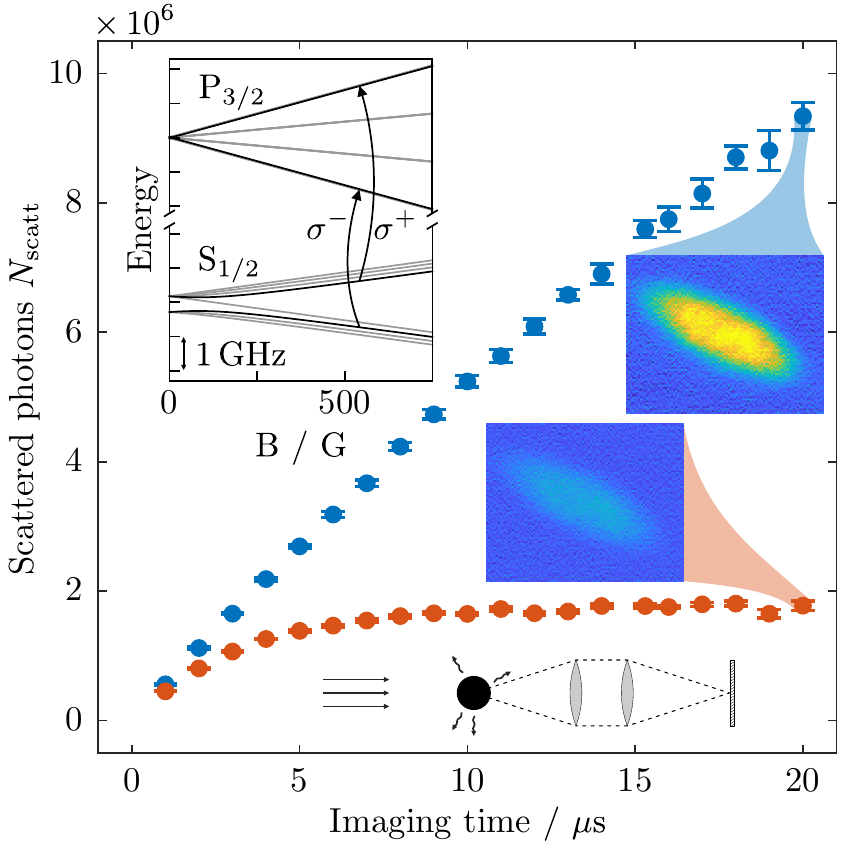}
    \caption{\textbf{Absorption imaging of a BEC of $^{39}$K.}
    The number of scattered photons $N_\mathrm{scatt}$ levels off within $\sim 5\,\mu \mathrm{s}$ when imaging the atoms with a single laser frequency ($\sigma^-$, red points). By adding a second frequency ($\sigma^+$),  the signal can be enhanced drastically (blue points). The difference is clearly visible in the absorption images of the atom cloud after $20\,\mu\mathrm{s}$ (same color scale used for both images; in the dark blue regions no photons are scattered). The upper inset shows the energy eigenstates of the ground state $\mathrm{S}_{1/2}$ and the excited state $\mathrm{P}_{3/2}$ hyperfine manifold as a function of the magnetic field $B$. The two imaging transitions are indicated with arrows. The lower inset shows a simplified schematic of the experimental setup. From left to right: laser light impinges on the atomic cloud, which is imaged via an objective and a secondary lens onto a CCD camera.
    }
    \label{fig1}
\end{figure}

\section{Experimental setting}

We exemplify the technique with a Bose-Einstein condensate (BEC) of $^{39}$K in the state which corresponds to $\lvert F,m_F\rangle =\lvert 1,-1\rangle$ at low magnetic fields. For our experiments we work at $550\,\mathrm{G}$, close to a broad Feshbach resonance.\cite{derricoFeshbachResonancesUltracold2007}
Figure 1 compares the absorption signal obtained with the improved absorption scheme (blue points) to the signal using only one laser frequency (red points). The latter results in a vanishing scattering after $\sim 5\,\mu\mathrm{s}$. With the addition of the second frequency, a drastic enhancement is achieved. In the experimental setup, each laser frequency is generated by a dedicated external cavity diode laser that is offset-locked\cite{offsetlock} to the cooling laser stabilized on the D2 line of $^{39}$K. After double-pass AOM paths for pulsing, both laser frequencies are coupled into the same single-mode optical fiber with orthogonal polarizations and pass through the same quarter-wave plate after the fiber. A CCD camera detects the total absorption signal (see Fig.~1, lower inset).
The number of scattered photons is estimated by
$N_{\mathrm{scatt}} = - \mathcal{G}\,(C_{\mathrm{f}}-C_{\mathrm{i}})$,
where $C_{\mathrm{f}}$ and $C_{\mathrm{i}}$ are the number of integrated counts on the CCD camera with and without atoms, respectively. The factor $\mathcal{G}$ includes the camera gain as well as a correction factor for the solid angle of the objective, reflection loss along the imaging beam path, and the quantum efficiency of the camera.

\begin{figure}
    \centering
    \includegraphics{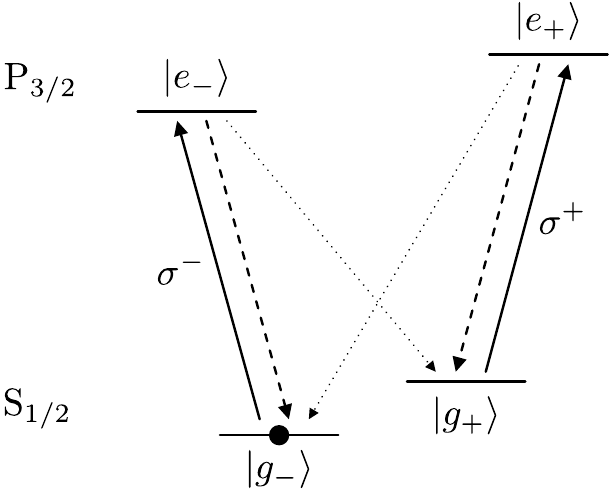}
    \caption{\textbf{Four-level scheme for imaging.}
    The imaging transition $\sigma^-$ transfers atoms from the initial state $\lvert g_-\rangle = \sqrt{p}\,\lvert -1/2, -1/2\rangle + \sqrt{1-p}\,\lvert 1/2, -3/2\rangle$ (black dot) to the excited state $\lvert e_-\rangle\simeq\lvert -3/2,-1/2\rangle$. Here, the states $\lvert m_J,m_I\rangle$ are the basis states of the electron's total angular momentum $\mathbf{J}$ and the nuclear spin $\mathbf{I}$.
    $p = 0.98$ at a magnetic field of $550\,\mathrm{G}$. Most of the atoms decay back to the initial state (dashed arrow), but a small leakage populates the state $\lvert g_+\rangle = \sqrt{p}\,\lvert 1/2, -3/2\rangle + \sqrt{1-p}\,\lvert -1/2, -1/2\rangle$ (dotted arrow). A second laser frequency drives the transition $\sigma^+$, which couples the state $\lvert g_+\rangle$ to the excited state $\lvert e_+\rangle\simeq\lvert 3/2, -3/2\rangle$ from where the atoms can decay back into $\lvert g_+\rangle$ and $\lvert g_-\rangle$ only. This results in a closed optical cycle when the excited states are sufficiently pure in quantum numbers $m_J,m_I$.
    }
    \label{fig2}
\end{figure}

\section{Principle of the method}

The upper inset of Fig.~1 shows the Breit-Rabi diagram of the $\mathrm{S}_{1/2}$ ground state and $\mathrm{P}_{3/2}$ excited state manifold, with the employed transitions depicted as arrows.
The relevant four-level scheme is depicted in Fig.~2. The atoms are initially prepared in $\lvert g_-\rangle \sim \lvert m_J,m_I\rangle = \lvert -1/2,-1/2\rangle$ with $m_J$ and $m_I$ denoting the magnetic quantum numbers of the electron's total angular momentum $\mathbf{J}$ and the nuclear spin $\mathbf{I}$, respectively.
Imaging at a single frequency involves a $\sigma^-$ transition to the state $\lvert e_-\rangle\simeq\lvert -3/2,-1/2\rangle$ in the $\mathrm{P}_{3/2}$ excited state manifold. The nearby states ($<15\,\mathrm{MHz}$) with the same $m_J$ are not addressed, since the nuclear spin quantum number $m_I$ is not changed by electric dipole transitions and the atomic eigenstates are pure up to $10^{-4}$ in the $\lvert m_J, m_I\rangle$ states. The excited state $\lvert e_-\rangle$ has a small leakage into a dark state $\lvert g_+\rangle\sim\lvert 1/2, -3/2\rangle$, which causes the quick saturation of the signal in Fig.~1.
Specifically, the two ground states read
\begin{align}
\lvert g_-\rangle = \sqrt{p}\,\lvert -1/2, -1/2\rangle + \sqrt{1-p}\,\lvert 1/2, -3/2\rangle , \nonumber \\
\lvert g_+\rangle = \sqrt{p}\,\lvert 1/2, -3/2\rangle + \sqrt{1-p}\,\lvert -1/2, -1/2\rangle ,\label{eq:groundstate}
\end{align}
with  $p \simeq 0.98$ at a field of $550\,\mathrm{G}$. As both ground states have an admixture of $\lvert -1/2,-1/2\rangle$, the excited state $\lvert e_-\rangle\simeq\lvert -3/2,-1/2\rangle$ can decay into both. The $2\,\%$ admixture is consistent with the observed time scale of $2.2\,\mu\mathrm{s}$, after which half of the atoms are transferred into the dark state. 

To enhance the signal we address the state $\lvert g_+\rangle$ with the second laser frequency. This $\sigma^+$ light couples $\lvert g_+\rangle$ to the excited state $\lvert e_+\rangle\simeq\lvert 3/2, -3/2\rangle$. It closes the optical cycle to good approximation and results in the effective four-level system shown in Fig.~2. During a typical $10\,\mu\mathrm{s}$ imaging pulse we expect to lose only $2\,\%$ of the atoms into the ground states $\sim\lvert -1/2,1/2\rangle$ and $\sim\lvert 1/2,-1/2\rangle$, which are not addressed by the imaging light. This results from the limited purity of the excited states in the $ \lvert m_J,m_I\rangle$ basis. Off-resonant coupling to other excited states is negligible for typical imaging intensities since the closest transitions are detuned by at least $350\,\mathrm{MHz}$.

\begin{figure}
    \centering
    \includegraphics{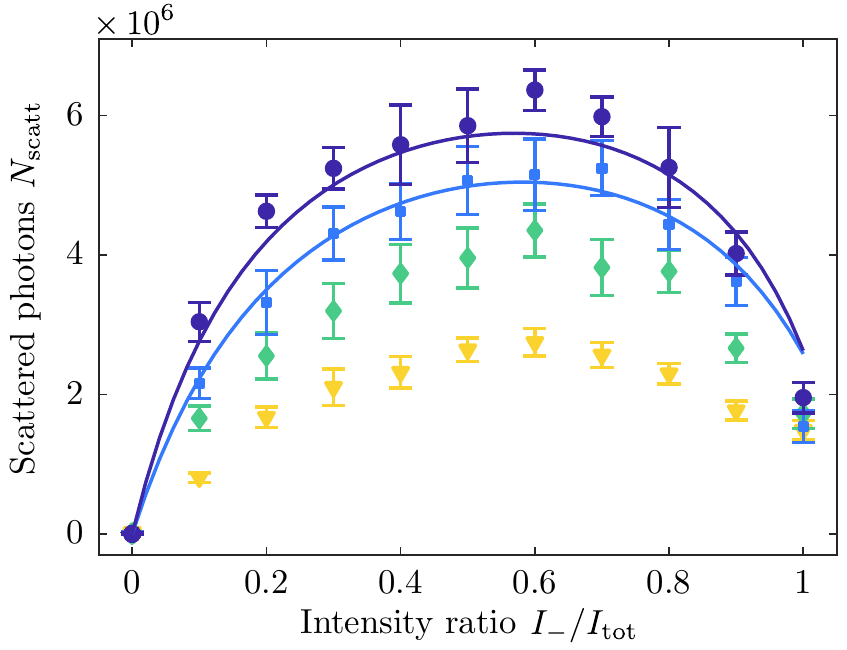}
    \caption{\textbf{Optimization of the intensity ratio.}
    The number of scattered photons is measured at different ratios $r=I_-/I_\mathrm{tot}$ and different total intensities $I_\mathrm{tot}=I_-+I_+$. The imaging pulse length is $10\,\mu\mathrm{s}$, and the data points correspond to total intensities of
    $23\,\mathrm{mW/cm^2}$ (triangle),
    $42\,\mathrm{mW/cm^2}$ (diamond),
    $60\,\mathrm{mW/cm^2}$ (square),
    $79\,\mathrm{mW/cm^2}$ (circle).
    For the highest intensities, the largest signal is found at $r\simeq0.5$.
    For decreasing light intensities this optimum slightly shifts to larger ratios as it takes longer to reach the steady state. For the highest intensities we can compare the data to numerical solutions of the optical Bloch equations for the four-level system, scaled by a global factor (solid curves).
    }
    \label{fig3}
\end{figure}

\section{Optimal intensity ratio}
 
We optimize the absorption signal by varying the ratio $r=I_-/I_\mathrm{tot}$ between the intensities $I_-$ and $I_+$ on the two imaging transitions $\sigma^-$ and $\sigma^+$, respectively. Here, the total imaging beam intensity $I_\mathrm{tot} = I_- + I_+$ is kept constant. Figure 3 shows the number of scattered photons $N_\mathrm{scatt}$ for different configurations and compares the results to numerical solutions of the optical Bloch equations for the four-level system (scaled by a constant factor). In the case without $\sigma^+$ light ($r=1$) the total signal is limited by the decay into the dark state. Imaging without $\sigma^-$ light ($r=0$) results in no signal, as the initial state of the atoms is not addressed by this light. For the highest imaging intensities the maximum number of scattered photons is obtained at $r\simeq 0.5$, as expected from the steady state solution.
For smaller intensities, the optimum is at higher ratios $r$. This results from the initial pumping dynamics starting in $\lvert g_-\rangle$ that are still relevant for the short imaging duration of $10\,\mu\mathrm{s}$.

\begin{figure}
    \centering
    \includegraphics[width=\columnwidth]{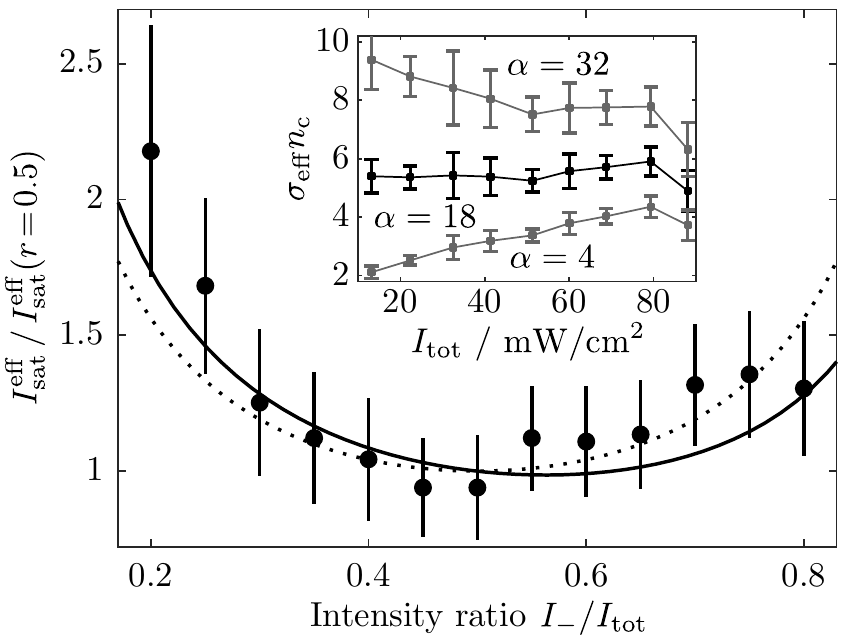}
    \caption{\textbf{Calibration of the imaging system.}
    For each ratio $r=I_-/I_\mathrm{tot}$ the effective saturation intensity $I_{\mathrm{sat}}^{\mathrm{eff}} = \alpha I_{\mathrm{sat}}$ is chosen such that the resulting atomic column density $n_c$ is invariant under changes of the imaging intensity $I_\mathrm{tot}$. The inset shows this procedure for $r=0.5$. The theoretical predictions obtained from the steady state solution and the numerical simulation of the dynamics are shown by the dashed and the solid curves, respectively. The experimental values are scaled by the mean of the three points around $r=0.5$, and the theoretical curves by their respective values at $r=0.5$. The error bars are estimated by bootstrap resampling.
    }
\end{figure}

\section{Calibration}

To obtain an accurate estimate of the atomic density, we calibrate the imaging system following the method presented in Reinaudi et~al.\cite{reinaudiStrongSaturationAbsorption2007} Each atom in the cloud is described as an effective two-level system, which includes saturation effects. One solves the resulting Beer-Lambert-type differential equation for resonant light and integrates along the direction of the imaging beam. This leads to the atomic column density
\begin{equation}
n_\mathrm{c}=\frac{1}{\sigma_\mathrm{eff}}\left[\ln \left(\frac{I_{\mathrm{i}}}{I_{\mathrm{f}}}\right)+\frac{I_{\mathrm{i}}-I_{\mathrm{f}}}{I_{\mathrm{sat}}^{\mathrm{eff}}}\right].
\end{equation}
Here, the final intensity $I_{\mathrm{f}}$ and the initial intensity $I_{\mathrm{i}}$ are the total intensities measured via the signal on the CCD camera with and without the presence of atoms, respectively.
$\sigma_\mathrm{eff}$ is the effective scattering cross-section and $I_{\mathrm{sat}}^{\mathrm{eff}}=\alpha I_{\mathrm{sat}}$ is the effective saturation intensity. The deviation from the bare saturation intensity $I_{\mathrm{sat}}$
of a single closed two-level optical cycle
captures effects of polarization, detuning fluctuations of the laser from atomic resonance, and optical pumping effects.  We estimate the effective saturation intensity by taking absorption images for a constant atom number with different total imaging intensities. 
$I_{\mathrm{sat}}^{\mathrm{eff}}$ is optimized such that the column density $n_\mathrm{c}$ is invariant under changes in intensity. As shown in the inset of Fig.~4, we find an optimum for $I_\mathrm{sat}^\mathrm{eff} = (18\pm 4) I_\mathrm{sat}$, where $I_\mathrm{sat}$ is the saturation intensity of a single transition. 
With the value of $I_\mathrm{sat}^\mathrm{eff}$ at hand an absolute atom number can be calibrated by comparing atomic density distributions with theoretical predictions \cite{TwoDYefsah} or the detection of atomic shot noise.\cite{ReadoutMussel}

To predict a value for  $I_{\mathrm{sat}}^{\mathrm{eff}}$ we scale the theoretical results for the scattering rate versus intensity of the four-level system to  the expectation for an effective two-level system. For the steady state the analytic solution reads
\begin{equation}
I_{\mathrm{sat}}^{\mathrm{eff}} (r) = \frac{I_\mathrm{sat}}{2r(1-r)}\;.
\end{equation}
We use that the coupled four-level system can be described by two two-level systems with equal $I_\mathrm{sat}$ which are only coupled to each other via the incoherent spontaneous decay of their excited states. Thus, no coherence is built up and the two subsystems can be described as being independent. In the steady state this leads to an imbalance in population of the two systems for $r\neq0.5$. In the case of $r=0.5$ the two populations are equal and the effective saturation intensity is twice the value of the single two-level system. We attribute the remaining deviation in the absolute value between experimental and theoretical $I_{\mathrm{sat}}^{\mathrm{eff}}$ mainly to instabilities of the imaging laser frequencies.

In Fig.~4 we show experimental results for the dependence of $I_{\mathrm{sat}}^{\mathrm{eff}}$ on the ratio $r$ and compare them to the analytic solution (dashed curve). At the largest and smallest ratios, deviations between experimental and analytic behavior arise due to the initial population dynamics of the four-level system.  These can be captured by a numerical simulation, as shown by the solid curve in Fig.~4. As before, the numerical results for the scattering rate versus total intensity are scaled to those of an effective two-level system. From $r\sim0.4$ to $0.6$ the effective saturation intensity varies only slightly, making the calibration of the column density robust against small changes of the imaging intensities.

\section{General perspectives}

Finally, we note that the imaging procedure can be generalized to all alkali-like atoms. The ground states can always be written as a superposition of maximally two $\lvert m_J,m_I\rangle$ states. This is a consequence of the the fact that the spin operator $F_z$ commutes with the Hamiltonian
\begin{equation}
H = a_{hf}/\hbar^2\ \mathbf{J}\cdot\mathbf{I} + \mu_B B_z/\hbar\ (g_J J_z + g_I I_z)
\end{equation}
of the ground state hyperfine manifolds. Here, $a_{hf}$ is the magnetic dipole constant and $g_J, g_I$ the electron and nuclear $g$-factors, respectively. This means that the z-projection $m_F = m_J + m_I$ of $\mathbf{F}$ is always a good quantum number. Since $J = 1/2$ for the ground states of all alkali atoms (i.e. $m_J=\pm1/2$) there are maximally two states with the same $m_F$. Except for the stretched states with maximal $\vert m_F \vert$, all states can be written in the form of Eq.~\ref{eq:groundstate}, and the imaging scheme can be applied.

\begin{acknowledgments}
We thank Selim Jochim, Martin Gärttner and Benedikt Erdmann for discussions. This work was supported by the DFG Collaborative Research Center SFB1225 (ISOQUANT), the ERC Advanced Grant Horizon 2020 EntangleGen (Project-ID 694561), and the Deutsche Forschungsgemeinschaft (DFG) under Germany's Excellence Strategy EXC-2181/1 - 390900948 (the Heidelberg STRUCTURES Excellence Cluster). M.~H.~acknowledges support from the Landesgraduiertenförderung Baden-Württemberg, C.~V. and N.~L.~acknowledge support from the Heidelberg graduate school for physics.
\end{acknowledgments}

\section*{Data Availability}

The data that support the findings of this study are available from the corresponding author upon reasonable request.

\end{document}